\documentclass[prl,aps,reprint,10pt,nofootinbib,preprintnumbers,superscriptaddress]{revtex4-2}
\pdfoutput=1
\usepackage{geometry}
\usepackage{color}
\usepackage{fullpage}
\usepackage[title]{appendix}
\usepackage{hyperref}
\hypersetup{colorlinks, citecolor=Blue, linkcolor=black, urlcolor=Blue}
\usepackage{amsfonts}
\usepackage{amsmath}
\usepackage{setspace}
\usepackage[usenames, dvipsnames]{xcolor}
\usepackage{tabularx}
\usepackage{empheq}
\usepackage{comment}
\usepackage{hepunits}
\usepackage{enumitem}
\usepackage{slashed}
\usepackage{braket}
\usepackage{tensor}
\usepackage{dsfont}

\definecolor{light-gray}{gray}{0.9}

\newcommand{\eps}{\epsilon}

\newcommand{\be}{\begin{eqnarray}}
\newcommand{\ee}{\end{eqnarray}}

\newcommand{\sD}{\slashed{D}}
\newcommand{\tr} {\mathrm{tr}}
\newcommand{\Tr} {\mathrm{Tr}}
\newcommand{\dc} {\mathcal{D}}
\newcommand{\dd} {\mathrm{d}}
\DeclareMathOperator{\g}{\sqrt{|\textit{g}|}}

\definecolor{lightviolet}{rgb}{0.75, 0.17, 0.8}

\definecolor{darkblue}{rgb}{0.0,0.0,0.6}
\definecolor{readableAB}{rgb}{0.7,0,0.7}
\definecolor{colorRTD}{rgb}{.2,.2,.7}

\begin{document}

\bigskip\

\title{Examining the Anomalous Nature of Chiral Effects in Thermodynamics}

\author{R\'emy Larue}
\affiliation{School of Physical Science and Technology, ShanghaiTech University,
393 Middle Huaxia Road, Shanghai 201210, China}
\author{J\'er\'emie~Quevillon}
\affiliation{Laboratoire d’Annecy-le-Vieux de Physique Th\'eorique,
CNRS – USMB, BP 110 Annecy-le-Vieux, 74941 Annecy, France}
\author{Diego Saviot}
\affiliation{Laboratoire d’Annecy-le-Vieux de Physique Th\'eorique,
CNRS – USMB, BP 110 Annecy-le-Vieux, 74941 Annecy, France}
\affiliation{Laboratoire de Physique Subatomique et de Cosmologie, Universit\'e Grenoble-Alpes, CNRS/IN2P3, Grenoble INP, 38000 Grenoble, France}

\begin{abstract}
Quantum anomalies give rise to novel transport phenomena, including the generation of a current in a relativistic fluid due to the presence of magnetic field or vorticity. We present an exclusive and direct computation of the chiral anomaly in the adiabatic regime within the path integral for a massless fermion on a generic electromagnetic and curved background, including local temperature and chemical potential. We identify new quantum-thermodynamical contributions to the breaking of axial symmetry which induce the Chiral Separation and Vortical Effects.
\end{abstract}

\maketitle


\textbf{\emph{Introduction.}}  --- The concept of quantum anomalies emerged in the late 1960s, first identified through the quantum breaking of the axial symmetry in quantum electrodynamics~\cite{Bell:1969ts,Adler:1969gk}. Since then, anomalies have become a cornerstone of research in theoretical physics. Surprisingly, the connection between the chiral anomaly and transport phenomena was recognized only recently, giving rise to effects such as the Chiral Magnetic Effect (CME)~\cite{Kharzeev:2008}, the Chiral Separation Effect (CSE)~\cite{Metlitski:2005pr} and the Chiral Vortical Effect (CVE)~\cite{Son_Surowka_2009,Kharzeev:2010gr}.

In condensed matter, the CME has been measured in semimetals~\cite{CME-Nature}; experimental evidence is also awaited and actively sought in heavy-ion collisions~\cite{Kharzeev_2024}; Weyl superfluids~\cite{Bevan:1997pqv,Volovik:2016rgb}; and important implications are expected in astrophysics and cosmology~\cite{Kamada:2022nyt}. These connections have established anomalous transport as a highly dynamic field of research.

For a Dirac fermion coupled to electromagnetism, with finite vector and axial chemical potentials $\mu$ and $\mu_5$ at temperature $T$, the vector and axial currents include
\begin{equation}
\begin{aligned}
    \langle j^\mu_{V}\rangle &\supset
    \frac{\mu_5}{2\pi^2} B^\mu
    +\frac{\mu \mu_5}{\pi^2}\Omega^\mu\;,\\
    \langle j^\mu_{5}\rangle &\supset
    \frac{\mu}{2\pi^2} B^\mu
    +\left(\frac{\mu^2+\mu_5^2}{2\pi^2}+\frac{T^2}{6}\right) \Omega^\mu\;,
    \nonumber
\end{aligned}
\end{equation}
where the CME (resp. CSE) is induced by the magnetic field $B^\mu$ in the vector (resp. axial) current, whereas the CVE is induced in both currents by the vorticity $\Omega^\mu = \frac{1}{2}\epsilon^{\mu\nu\rho\sigma}u_\nu \partial_\rho u_\sigma$, where $u_\mu$ is the fluid 4-velocity. 

More generally, the decomposition of macroscopic currents, including the CME, CSE, and CVE, has been formulated within the framework of the hydrodynamic effective action~\cite{Bhattacharya:2011tra, Banerjee:2012iz, Kovtun:2018dvd}. This `bottom-up' construction is independent from a microscopic theory and thus lists all the possible contributions to the currents, up to unknown transport coefficients. These coefficients can be accessed via a hydrodynamics approach~\cite{Son_Surowka_2009, Neiman:2010zi}, or using linear response theory through the so-called Kubo formulae~\cite{Amado_Landsteiner_Pena-Benitez_2011, Landsteiner:2011cp,Chernodub:2021nff}, as well as other methods~\cite{Amado_Landsteiner_Pena-Benitez_2011,Chen_Son_Stephanov_2015,Liu:2018xip}.

In the hydrodynamic picture, a connection between the chiral anomaly and the chiral effects involving the chemical potential was initially obtained in~\cite{Son_Surowka_2009}. However, such a link remains elusive from the microscopic point of view as the anomaly is injected and not computed within the same framework. 
It was proposed in~\cite{Landsteiner:2011cp}, and subsequently explored in~\cite{Stone:2018zel, Landsteiner:2023,Prokhorov:2022udo, Flachi_2018}, that the temperature dependence of the CVE is induced by, or at least intimately linked to, the axial-gravitational anomaly.
On the other hand, it is well-established that the chiral and axial-gravitational anomalies are independent from a temperature and chemical-potential, which seems to contradict these connections. The temperature-independence was shown in~\cite{Itoyama:1982up, Reuter_Dittrich_1985, Chaturvedi_Gupte_Srinivasan_1985,Das:1987yb, Treml:1990ce} for the chiral anomaly, and in~\cite{Boschi-Filho_Natividade_1992,Corianò_Cretì_Lionetti_Tommasi_2024} for the axial-gravitational anomaly, whereas the independence from the chemical potential was investigated in~\cite{Sisakian:1997cp}.

In this paper, we focus on the CVE and CSE for a massless Dirac fermion and investigate their relation to the chiral anomaly in the presence of local temperature and chemical potential, from a `top-down' approach.
In the first Section, we introduce the imaginary-time path integral representation of the local equilibrium partition function with background curvature and electromagnetic field, and discuss global equilibrium conditions.
In the second Section, we then access the anomaly by evaluating the Jacobian of the chiral transformation as a ratio of path integrals.
This also yields the quantum-statistical expectation value of the axial current itself and generalises the CVE and CSE in gravity and for any fluid velocity. We end the paper with a discussion and conclusions, and show the consistency of our new terms with the usual anomaly equation.\\

\textbf{\emph{Thermodynamics in Curved Spacetime.}} --- Our goal is to study a spin-1/2 Dirac fermion with mass $m$ coupled to a  gravitational and electromagnetic background, at local equilibrium.\\

\paragraph{Temperature in imaginary-time formalism}  ---
We first consider a flat spacetime, where the local equilibrium is described by the distribution that minimises the information entropy. At constant temperature $T_0=1/\beta_0$ and chemical potential $\mu_0(\vec x)$ associated with the vector current $j^\mu_V=\bar\psi\gamma^\mu\psi$, the partition function reads~\cite{LANDAU_LIFSHITZ_1987}
\begin{equation}
    Z=\Tr\,e^{-\beta_0(H-\int d^3 x \mu_0(\vec x) j^t_V)}\;.
\end{equation}
Using the imaginary-time formalism it can be recast as a path integral at a fixed real-time $t_R$~\cite{LeBellac_1996,Hongo_2017,Hong:2022nss} 
\begin{equation}
    Z[t_R]=\int_{b.c}\dc\bar\psi\dc\psi e^{-\int_0^{\beta_0}\dd t\int\dd^3x\,\mathcal{L}}\;, \label{eq:W}
\end{equation}
with $\mathcal L=\bar\psi\left(i\gamma^\mu(\partial_\mu+i\mathcal{V}_\mu)-m\right)\psi$ and $\mathcal{V}_\mu= V_\mu + \mu_0 \delta_\mu^t$ where $V_\mu$ is the vector gauge field. The boundary conditions are anti-periodic in imaginary-time  $b.c = \{ \psi(i\beta_0,\vec{x}) = -\psi(0,\vec{x}),\;\bar\psi(i\beta_0,\vec{x}) = -\bar\psi(0,\vec{x}) \}$.  Importantly in the imaginary-time formalism for local equilibrium, the path integral variables depend only on the imaginary-time $t$ and therefore ${\cal L}\supset\partial_t$, whereas all the background fields and thermodynamic variables depend only on the real-time $t_R$~\cite{Hongo_2017,Hong:2022nss}.  As a result, the expectation values are obtained in the adiabatic regime where $\partial_{t_R}\langle \mathcal{O}\rangle\simeq\langle\partial_{t_R}\mathcal{O}\rangle\simeq 0$. In the remaining of the paper we omit the $t_R$ dependence for simplicity.
\\

\paragraph{Electro-chemical potential}  ---
The decomposition of the electro-chemical potential $\mathcal{V}_t=V_t+\mu_0$ into an electrostatic potential $V_t$ and an internal chemical potential $\mu_0$ is arbitrary~\cite{Newman_Balsara_2021}. As it ultimately defines the thermodynamic parameter that determines the vector charge density, we will refer to it as $\mu_{ec}$ in the following. 
To distinguish between the chemical and electrostatic components of the electro-chemical potential, a reference point must be introduced. A natural choice is to redefine the electrostatic potential such that it vanishes when the electric field does: $U_{elec}=V_t-V_t|_{\vec{E}=\vec{0}}$ with $V_t|_{\vec{E}=\vec{0}}=const$. In this case, the chemical potential is defined as $\mu\equiv\mu_{ec}-U_{elec}=\mu_0+\left.V_t\right|_{\vec{E}=0}$. Although the separation $\mu_{ec}=\mu+U_{elec}$ depends on the reference point, the gradient $\partial_\nu \mu_{ec} = \partial_\nu \mu +E_\nu$ does not.\\

\paragraph{Curved spacetime}  ---
In a curved spacetime admitting a foliation along the time direction $t^\mu(x)$, the notion of equilibrium requires the introduction of the 4-temperature $\beta^\mu=u^\mu/T$ with $u^\mu$ the  fluid 4-velocity with $u^2=1$ and $T(x)$ the local temperature. The local equilibrium partition function reads \cite{Weldon:1982aq, Becattini:2012tc,Hongo_2017} 
\begin{equation}
    Z = \Tr \,e^{-\int d\Sigma \,n_\mu(T^{\mu\nu}\beta_\nu-\zeta_0\, j^\mu_V)}\;,\label{eq:Zlocaleq}
\end{equation}
where $n_\mu(x)$ is the unit vector orthogonal to the hypersurfaces of constant time, and $\zeta_0=\mu_0(x)/T(x)$.

Without loss of generality, we may choose the coordinate system $\{x^\mu\}=\{t,\vec x\}$ such that $t^\mu=\delta^\mu_t$ and $u^\mu=t^\mu/\sqrt{|t^2|}$~\cite{Hongo_2017}. This choice may come at the expense of having a more complicated metric, which is not a difficulty since we will keep it as generic as possible. In particular, with this choice, the local temperature is $T(x)=T_0/\sqrt{|g_{tt}|}$ 
and the normal to the time-slices is $n_\mu=\delta_\mu^t/\sqrt{|g^{tt}|}$.~\footnote{Even though $\sqrt{g_{tt}} T=const$ due to our choice of coordinates, the temperature is not at global equilibrium, i.e $\beta^\mu$ is not Killing.}
The electro-chemical potential is $\mu_{ec}=\mu+U_{elec}/\sqrt{|g_{tt}|}$ and the electric field is defined as $E_\mu\equiv F_{\mu\nu}u^\nu=(\partial_\mu U_{elec})/\sqrt{|g_{tt}|}$.

The local equilibrium partition function can be recast as the path integral Eq.~\eqref{eq:W} with the curved space-time Lagrangian~\cite{Hongo_2017}
\begin{equation}
\hspace{-0.2cm}\mathcal L=\bar\psi\left(i\gamma^\mu(\nabla_\mu+\omega_\mu+i\mathcal{V}_\mu)-m\right)\psi,\; t^\mu\mathcal{V}_\mu=\sqrt{g_{tt}}\mu_{ec} \;.\label{eq:Lagr}
\end{equation}
We have $\gamma^\mu=\tensor{E}{_A^\mu}\gamma^A$ with $\tensor{E}{^A_\mu}$ the vierbein fields, $\omega_\mu$ is related to the spin-connection by $\omega_\mu=\frac{1}{8}[\gamma^A,\gamma^B]\omega_{\mu,AB}$ and $\nabla$ is the diffeomorphism covariant derivative. Note that the spacetime measure $\g$ was absorbed in the path integral variables so as to obtain a diffeomorphism-invariant path integral measure~\cite{Fujikawa:1980rc,Toms:1986sh,Fujikawa:2004cx}.
\\

\paragraph{Global equilibrium}  ---
The global equilibrium  conditions follow from requiring that the exponent of the partition function Eq.~\eqref{eq:Zlocaleq} is the same for each time-slice. A set of sufficient conditions are~\cite{Israel:1979wp,Chrobok:2006rr,Becattini:2012tc}~\footnote{Less restrictive conditions can be found~\cite{Lima:2019brf}.}
\begin{itemize}
    \item Temperature: $\beta_\mu$ is Killing, i.e $\nabla_\mu\beta_\nu+\nabla_\nu\beta_\mu=\frac{1}{T_0}\partial_{t_R} g_{\mu\nu}=0$, that is to say the metric is independent from the real-time. In that case, $T(x)=1/\sqrt{\beta^2}$ is the global equilibrium temperature~\cite{Tolman:1930zza,Tolman:1930ona,Luttinger:1964zz}.
    \item Electro-chemical potential: $\mathcal{E}_\mu\equiv T\partial_\mu\frac{\mu_{ec}}{T}=0$ or equivalently $T\partial_\mu\frac{\mu}{T}+E_\mu=0$. Close to global equilibrium, $T\partial\frac{\mu}{T}$ is the electric field created by the distribution of charges in the medium, and $\mathcal{E}$ is therefore the total electric field.
\end{itemize}
Note that we only apply these conditions at the end of our computations.
\\

\paragraph{Fluid velocity within the metric}  ---
Beyond applications to gravity, the curvature of spacetime can also describe a fermion fluid in motion. For example, consider the flat metric $g_{\mu\nu}=\eta_{\mu\nu}=\mathrm{diag}(1,-1,-1,-1)$ and a fluid with 4-velocity $u^\mu\equiv t^\mu/\sqrt{|t^2|}=\frac{1}{\sqrt{1-\vec v^2}}(1, -\vec v)$. If the fluid has a vanishing material derivative {$\partial_t\vec v-v^i\partial_i\vec v=0$}, then the change of coordinate to the fluid rest frame is simply achieved by $t'=t$ and $x'^i=x^i+v^i t$. In the rest frame, the time-direction is given by $t^\mu=\delta^\mu_t$ and the 4-velocity $u^\mu=\frac{1}{\sqrt{1-\vec v^2}}(1, \vec 0)$. As mentioned above, the metric is now more complicated and reads
\begin{equation}
g'_{\mu\nu}=\begin{pmatrix}
1-\vec v\,^2 & \vec v\\
\vec v\,^T & -\mathds{1}_3
\end{pmatrix}\;.\label{eq:metricfluid}
\end{equation}
Note that in imaginary-time the fluid is described by a complex metric.\footnote{See Refs.~\cite{PhysRevLett.129.242002,Chen:2024tkr} for interesting discussions about rotating fluids with imaginary fluid velocity.} Since we kept the metric generic, our partition function~\eqref{eq:Lagr} can  describe a fluid with any velocity in a gravitational background, which is in general more complicated than~\eqref{eq:metricfluid}.\\

\textbf{\emph{Chiral Anomaly and Chiral Effects.}} --- We consider the $U(1)_A$ transformation  of infinitesimal parameter $\theta(t_R,\vec x)$ with $\theta(|\vec x|\to\infty)=0$
\begin{equation}
\psi'=e^{i\theta \gamma_5}\psi\;,\quad\quad \bar\psi'=\bar\psi e^{i\theta\gamma_5}\;.\label{eq:U(1)A}
\end{equation}
The partition function Eq.~\eqref{eq:W} with Lagrangian Eq.~\eqref{eq:Lagr} reads~\footnote{Note that the functional determinant and traces are over functions that respect the boundary conditions $b.c$.}
\begin{equation}
Z=\det\left[i\sD-\gamma^t\sqrt{g_{tt}}\mu_{ec}-m\right]\;,\quad D_\mu\equiv\nabla_\mu+\omega_\mu+i\delta^i_\mu V_i\;,\nonumber
\end{equation}
where we separated $\mu_{ec}$ from the covariant derivative for convenience.
It is invariant under relabeling its integration variables as Eq.~\eqref{eq:U(1)A} which produces a Jacobian $J[\theta]$~\cite{Fujikawa:1979ay}
\begin{align}
&Z=J[\theta]\det\left[i\sD-\gamma^t\sqrt{g_{tt}}\mu_{ec}-m-(\slashed\partial\theta)\gamma_5-2im\theta\gamma_5\right]\nonumber\\
&\text{with }\;J[\theta]\equiv \exp\left(-\int_0^{\beta_0} \dd t\int\dd^3x \,\sqrt{g}\,\theta\mathcal{A}\right)\;.\label{eq:Jac}
\end{align}
The anomaly can thus be expressed as a ratio of functional determinants~\cite{Filoche:2022dxl}, and in particular we obtain
\begin{equation}
\log J[\theta]=\Tr\left[\Big((\slashed\partial\theta)\gamma_5+2im\theta\gamma_5\Big)\frac{1}{i\sD-\gamma^t\sqrt{g_{tt}}\mu_{ec}-m}\right]\;.\nonumber
\end{equation}
At zero temperature, one cannot straightforwardly take the $m=0$ limit due to the presence of infrared (IR) divergences, and the term proportional to $2im\theta\gamma_5$ is necessary to obtain the anomaly~\cite{Filoche:2022dxl}. However, at finite temperature, the massless limit is well-defined since the temperature acts as an IR regulator as we will see below, and the $2im\theta\gamma_5$-term can safely be dropped. In that case, we recover that the anomaly (in the adiabatic regime where the time-derivatives vanish) is the spatial divergence of the expectation value of the axial current $\mathcal{A}=\langle\nabla_i j^i_5\rangle$.

We evaluate the Jacobian using the Covariant Derivative Expansion (CDE) in curved spacetime (e.g~\cite{Larue:2023uyv}) adapted to finite temperature and imaginary-time, which is organised as a weak field expansion compared to $\mathrm{Max}\{\mu_{ec},T\}$ with Euclidean metric
\begin{align}
&\log J[\theta]\underset{m=0}{=}\int_0^{\beta_0}\!\!\!\!\dd t\int\!\!\dd^3x\,\sqrt{g}(\partial_\mu\theta)\langle j^\mu_5\rangle\label{eq:anomCDE}\\
&=-\int_0^{\beta_0}\!\!\!\!\dd t\int\!\!\dd^3x(\partial_\mu\theta)\,\frac{1}{\beta_0}\sum_{n\in\mathds Z}\int\frac{\dd^3 q}{(2\pi)^3}\tr\,\gamma^\mu\gamma_5\sum_{k\in\mathds{N}}\left[\Delta i\sD\right]^k\Delta\;,\nonumber
\end{align}
where the propagator $\Delta$ involves the fermionic Matsubara frequencies $\Omega_n=(2n+1)\pi T_0-i\sqrt{g_{tt}}\mu_{ec}(x)$,
\begin{equation}
\Delta(q)=\frac{\gamma^t \Omega_n+\gamma^i q_i}{g^{tt}\Omega_n^2+2g^{ti}\Omega_n q_i+g^{ij}q_i q_j}\;.\label{eq:Delta}
\end{equation}
Note that we chose to insert the electro-chemical potential in the Matsubara frequencies, although we could equivalently have kept it in the covariant derivative\footnote{
Keeping $\mu_{ec}$ in the propagator allows for a non-perturbative treatment rather than considering it to be small with respect to the temperature. It is a resummation of the perturbative case, as happens in the study of strong-fields, see \cite{Franchino-Vinas:2025eze}.}.

To compute the Matsubara sums it is convenient to change variables to $p_i=q_i-\Omega_n g_{ti}/g_{tt}$ which yields
\begin{equation}
\Delta(p)=
\frac{u_\mu\gamma^\mu\tilde\Omega_n+\gamma^i p_i}{\tilde\Omega_n^2+g^{ij}p_i p_j}\;,\quad \tilde\Omega_n=(2n+1)\pi T(x)-i\mu_{ec}(x)\;.
\end{equation}
From this expression, it is clear that the temperature prevents the denominator from vanishing, hence acts as an IR regulator.
Our conventions are defined in App.~\ref{app:conventions}. Details about the CDE are deferred to App.~\ref{app:CDE}.\\

\paragraph{Chiral Vortical Effect}  ---

The leading order contribution to the chiral anomaly occurs at $k=1$ in Eq.~\eqref{eq:anomCDE}. The expansion is carried out keeping in mind that the covariant derivative does not commute with the propagators due to the presence of the metric and the electro-chemical potential.
Nonetheless, the only contribution that is non-vanishing under the Dirac trace comes from the spin-connection
\begin{equation}
\left.\langle j^\mu_5\rangle\right|_{k=1}=\frac{-i}{\sqrt{g}}\left(\mathcal{I}^2[q^2]\right)_{\alpha\beta}\tr\,\gamma^\mu\gamma_5\gamma^\alpha\gamma^\nu\gamma^\beta\omega_\nu\;,
\end{equation}
with the master integrals defined in App.~\ref{app:MasterI}. These integrals can be straightforwardly evaluated in the massless limit and we find in Lorentzian signature the CVE~\footnote{The contribution from the vacuum ($T,\mu_{ec}=0$) vanishes when regularised such that the vector symmetry is preserved, see App.~\ref{app:CVE}.}
\begin{align}\label{eq:j5CVEnoeq}
\left.\langle j^\mu_5 \rangle\right|_{k=1}&=\left(\frac{\mu_{ec}^2}{2\pi^2}+\frac{ T^2}{6}\right)\Theta^\mu\;,
\end{align}
where~\footnote{Note that $\Theta$ is expressed in terms of the spin-connection in the inertial frame where $u^\mu\propto(1,\vec 0)$ only. This is not a frame-independent expression. However $\Theta^\mu=\frac{1}{2}\epsilon^{\mu\nu\rho\sigma}u_\nu\partial_\rho u_\sigma$ remains valid in any frame.}
\begin{equation}\label{eq:Theta}
    \Theta^\mu=-\frac{1}{2}\epsilon^{\mu\nu\rho\sigma}u_\nu u^\lambda \omega_{\lambda,\rho\sigma}=\frac{1}{2}\epsilon^{\mu\nu\rho\sigma}u_\nu \partial_\rho u_\sigma\;,
\end{equation}
is the vorticity. Note that this expression is valid for any fluid velocity and gravitational background in the adiabatic regime. It is only in flat spacetime and for a fluid velocity with vanishing material derivative that the metric takes the simple form~\eqref{eq:metricfluid} and $u_i$ is proportional to the fluid velocity $v_i$.

Although the CVE is not related to the usual axial-gravitational anomaly ($\propto R\tilde R$) as discussed in~\cite{Stone:2018zel, Landsteiner:2023}, it is still a mixed anomaly in the sense that is it induced by a gravitational background (the spin-connection here) in the axial current.

To obtain the contribution of the CVE to the anomaly, we integrate by parts the derivative on $\theta$ in Eq.~\eqref{eq:anomCDE}. Since the time derivative is with respect to the imaginary-time and that the background fields depend only on the real-time, we have in particular $\nabla_\mu(T^2\Theta^\mu)=0$ (see App.~\ref{app:CVE}). Finally, we obtain
\begin{align}
\begin{split}\label{eq:ACVESec2}
\mathcal{A}_{\mathrm{CVE}}=&\frac{\mu_{ec}}{\pi^2}\vec {\cal E}\cdot\vec\Theta=\frac{\mu_{ec}}{\pi^2}\left(T \partial_i\frac{\mu}{T}+E_i\right)\Theta^i\;.\\
\end{split}
\end{align}
To the best of our knowledge, this contribution of the CVE to the anomaly is an important and new result.
Importantly, it can be recast as a Pontryagin density for the vector field $U_\mu=\mu_{ec}u_\mu$.~\footnote{This was also pointed out in~\cite{Nissinen:2021gke} for a constant chemical potential and in flat spacetime
\begin{equation}\label{eq:CVEPontryagin}
\mathcal{A}_{\text{CVE}}=\frac{1}{2\pi^2}F_U\tilde F_U\;,
\end{equation}
with $F_{U,\mu\nu}=\nabla_\mu U_\nu-\nabla_\nu U_\mu$ and $\tilde F_U^{\mu\nu}=\frac{1}{2}\epsilon^{\mu\nu\rho\sigma}F_{U,\rho\sigma}$. This reflects the fact that the density of charges $\mu_{ec}$ in the fluid moving with 4-velocity $u_\mu$ is an electric current from the point of view of the observer. If the fluid is rotating, the observer will therefore measure a magnetic field  $\vec B_{\Theta}=2\mu_{ec}\vec\Theta$.
Note that the $T^2\Theta$-term can also be recast as a Pontryagin density for $U'_\mu=T u_\mu$, although it does not contribute to the anomaly in the adiabatic regime and is thus beyond the scope of this paper.
}

Lastly, when the global equilibrium conditions 
are enforced, one finds that the whole contribution of the CVE to the anomaly vanishes
\begin{equation}
\mathcal{\mathcal{A}_{\mathrm{CVE}}}\underset{\text{GEQ}}{=}0\;.   
\end{equation}
More details about the CVE calculation and the CVE at global equilibrium are deferred to App.~\ref{app:CVE}.\\

\paragraph{Chiral Separation Effect}  ---
The next-to-leading order contribution occurs at $k=2$ in Eq.~\eqref{eq:anomCDE}. For simplicity, we omit terms that involve more than one derivative of the fields and thermodynamic quantities.
As detailed in App.~\ref{app:CSE}, the computation involves both $\left(\mathcal{I}^3[q^3]\right)_{\alpha\beta\gamma}$, which is finite since its vacuum contribution vanishes identically, and $\left(\mathcal{I}^4[q^4]\right)_{\alpha\beta\gamma\delta}$ which is ultraviolet (UV) divergent. We regularise it using dimensional regularisation
\begin{equation}
\frac{1}{\beta_0}\sum_n\int\frac{\dd^3q}{(2\pi)^3}\longrightarrow\frac{1}{\beta_0}\sum_n\int\frac{\dd^dq}{(2\pi)^d}\;,\quad d=3-\epsilon\;.\nonumber
\end{equation}
Note that the Dirac matrices are also extended to $d$ dimensions, and we use the Breitenlohner-Maison-’t Hooft-Veltman (BMHV) scheme~\cite{tHooft:1972tcz,Breitenlohner:1977hr} for $\gamma_5$ which preserves the vector symmetry (see \textit{e.g}~\cite{Filoche:2022dxl}).

Importantly, all divergences cancel when all contributions are taken into account. Nonetheless, finite terms that arise from UV poles $1/\epsilon$ mutiplied by an $\epsilon$ from the Dirac trace contribute to the final result. These terms are necessary to account for and ensure the gauge-invariance of the result. We thus obtain in Lorentzian signature
\begin{equation}
\left.\langle j^\mu_5\rangle\right|_{k=2}=\frac{1}{2\pi^2}\mu_{ec} B^\mu+\mathcal{O}(\partial^2(g,\mu_{ec},T))\label{eq:CSE}
\end{equation}
where the magnetic field is $B^\mu=\frac{1}{2}\epsilon^{\mu\nu\rho\sigma}u_\nu F_{\rho\sigma}$ with $F_{ij}=\partial_i V_j-\partial_j V_i$ and $F_{ti}=0$ since we moved $V_t$ in the electro-chemical potential and $\partial_t V_i=0$.
We recognise the CSE now generalised to include the effects of a background curvature, and background electric field which is encompassed in the electro-chemical potential.

After taking the divergence and applying $\delta/\delta\theta(x)$, we find the contribution of the CSE to the anomaly
\begin{equation}\label{eq:A-CSE}
\mathcal{A}_{\mathrm{CSE}}=\frac{1}{2\pi^2}\vec{\cal E}\cdot \vec B=\frac{1}{2\pi^2}\left(T\partial_i\frac{\mu}{T}+E_i\right) B^i
\end{equation}
Again, at global equilibrium the contribution of the CSE to the anomaly vanishes
\begin{equation}
    \mathcal{A}_{\mathrm{CSE}}\underset{\text{GEQ}}{=}0\;.
\end{equation}
Note that the contribution from the CSE to the anomaly is equal to the Pontryagin density for the gauge field $\mathcal V_\mu=\sqrt{g_{tt}}\mu_{ec}\delta^t_\mu+V_i\delta^i_\mu$
\begin{equation}
\mathcal{A}_{\mathrm{CSE}}=\frac{1}{2\pi^2}\vec {\cal E}\cdot \vec B=\frac{1}{8\pi^2}F_{\cal V}\tilde{F}_{\cal V}\;.\label{eq:ACSEPontryagin}
\end{equation}
Details about the derivation of the CSE are deferred to App.~\ref{app:CSE}.\\

\vspace{3pt}

\textbf{\emph{Discussion and conclusions.}} ---
In this paper, we compute for the first time the chiral
anomaly with  generic background curvature (i.e gravity and/or
fluid velocity) and electromagnetic field,
at local equilibrium with local temperature and electro-chemical potential, and in the adiabatic regime where the time-derivatives vanish.
We define the anomaly as the symmetry breaking which is not classical (i.e the Jacobian Eq.~\eqref{eq:Jac}),  arising from the quantum-statistical distribution. 
As a consequence, we do not only obtain the usual vacuum contribution to the  anomaly, but also the local thermodynamic parameters which is to our knowledge a new result with clear phenomenological applications. 
It is given by
\begin{equation}
\mathcal{A}=\langle\nabla_i j^i_5\rangle=\frac{1}{2\pi^2}\left(T\vec\partial\frac{\mu_{ec}}{T}\right)\cdot\left(2\mu_{ec}\vec \Theta+\vec B\right)\;,\label{eq:AnoeqConclusion}
\end{equation}
up to third derivatives of the fields and thermodynamic variables, with $\mu_{ec}$ the electro-chemical potential,  $\Theta^\mu=\frac{1}{2}\epsilon^{\mu\nu\rho\sigma}u_\nu\partial_\rho u_\sigma$ the vorticity and $\vec B$ the external magnetic field.

As a corollary result of this paper, we also obtain the generalisation of the CVE Eq.~\eqref{eq:j5CVEnoeq} and CSE Eq.~\eqref{eq:CSE} at local equilibrium with electromagnetic and curved background in the adiabatic regime.
One of the main advantages of our method of computation is that the link between the anomaly and the chiral effects is clearly established.

Eq.~\eqref{eq:AnoeqConclusion} differs from the anomaly in a simple electromagnetic background. The first main difference is the dependence on the gradient of chemical potential. Its presence is however rather intuitive and reflects the fact that the distribution of charges in the medium generates an electric field proportional to $T\partial\frac{\mu}{T}$. The anomaly is therefore sensitive to the total electric field $\vec{\cal E}=T\vec\partial\frac{\mu_{ec}}{T}=T\vec\partial\frac{\mu}{T}+\vec E$, including the one created by the macroscopic distribution of charges.

Additionally, the anomaly contains a term proportional to the vorticity $\Theta^\mu$. In the inertial frame, it arises from the spin-connection~\eqref{eq:Theta} and appears as a consequence of the separation of space and time at finite temperature. As a result it does not appear in the Fujikawa~\cite{Fujikawa:1979ay} or Leutwyler~\cite{Leutwyler:1984de,Leutwyler:1985ar,Leutwyler:1985em} procedures which restore Lorentz invariance (see App.~\ref{app:Leutwyler}). As shown in Eq.~\eqref{eq:CVEPontryagin} it can be expressed in terms of a Pontryagin density for the vector field $U_\mu=\mu_{ec}u_\mu$, which implies that $\vec B_{\Theta}=2\mu_{ec}\vec\Theta$ is in fact the magnetic field generated by the rotation of charges in the fluid as seen by the observer.

We emphasise on a potential confusion that can be found in the literature: our direct computation reveals that the electric and magnetic fields created by the distribution and motion of charges in the fluid enter the anomaly on the same footing as the external electric and magnetic fields.
In particular, in the bottom-up approach originating from~\cite{Son_Surowka_2009}, and retaken across the literature~\cite{Neiman:2010zi, Yang:2022ksq, Hidaka:2017auj}, the  anomaly equation should not be matched onto $\frac{1}{2\pi^2}\vec E\cdot \vec B$ but onto
\begin{equation}
\mathcal{A}=\frac{1}{2\pi^2}\vec {\cal E}\cdot\vec{\cal B}\;,
\end{equation}
where $\vec{\cal E}=T\vec\partial\frac{\mu_{ec}}{T}=T\vec\partial\frac{\mu}{T}+\vec E$ and $\vec{\cal B}=\vec B+2\mu_{ec}\vec\Theta$ are the total electric and magnetic fields as seen by the observer. An important consequence of this equation is that the CVE and CSE are in fact similar effects and are both generated by the chiral anomaly. The CSE is created by a non-uniform distribution of charges and an external electric in the presence of an external magnetic field, whereas the CVE is created by the rotation of a charged fluid in the presence of an electric field.

Finally, another important result of this paper is the vanishing of the chiral anomaly at global equilibrium
\begin{equation}
    \mathcal{A}\underset{\text{GEQ}}{=}0\;,
\end{equation}
which simply results from the vanishing of the total electric field $\vec{\cal E}$ at global equilibrium.

Another point raised in the literature is the relation between temperature dependence of the CVE and the mixed axial-gravitational anomaly $\propto R\tilde R$. This link is only realised in a specific setup where a black hole is introduced at infinity to obtain a geometric source for the temperature~\cite{Stone:2018zel, Landsteiner:2023}, although $R\tilde R$ is vanishing where the CVE is present. 
Eq.~\eqref{eq:j5CVEnoeq} demonstrates that the whole CVE indeed arises from a new mixed axial-gravitational term that involves the spin-connection. Incidentally, this solves the mismatch of power counting~\cite{Landsteiner:2023} between the four derivatives of the metric in $R\tilde R$ and the only two derivatives in the divergence of the CVE.

Finally, it would be interesting to investigate a non-adiabatic regime, since by naively taking the total divergence of the CVE and CSE including the time-derivative, one obtains additional contributions depending on the temperature gradient and the acceleration of the fluid $a_\mu=u^\lambda\nabla_\lambda u_\mu$~\footnote{Note that we have access to $\nabla_\mu\langle j^\mu_5\rangle$ and not $\langle\nabla_\mu j^\mu_5\rangle$, although the anomaly is the second one. The reason is that the (real-)time derivative and the expectation value do not commute for a non-stationary distribution.}
\begin{align}
\begin{split}
\nabla_\mu \langle j^\mu_5\rangle&=\frac{T^2}{6}\left(\nabla_\mu\Theta^\mu+2\Theta^\mu\frac{\partial_\mu T}{T}\right)\\
&+\frac{\mu_{ec}}{2\pi^2}\left(\frac{\partial_\mu T}{T}-a_\mu \right)B^\mu+\frac{1}{2\pi^2}\vec{\cal E}\cdot \vec {\cal B}\;.
\end{split}
\end{align}
In particular, a non-adiabatic regime would allow to directly link the $T^2\Theta$-term of the CVE to the anomaly.  
This would however require the use of a real-time approach such as the Schwinger-Keldysh formalism~\cite{Schwinger:1960qe,Keldysh:1964ud} which is out of the scope of this paper.

Our approach should be well-suited to investigate the transport effects with axial background (e.g the CME), and should offer novel insights. Since it involves additional subtleties regarding the introduction of an axial chemical potential and the treatment of $\gamma_5$ in dimensional regularisation, we postpone the discussion for future work~\cite{Larue:2026cjx}. Our findings about the new contribution to the chiral anomaly at local equilibrium could have interesting implications regarding the problem of matter anti-matter asymmetry in the universe, e.g baryogenesis, and also phenomenology of axions~\cite{LMQS:2025}.\\

\textit{Acknowledgements.} --- We thank M. Chernodub, C. Delaunay, A. Grushin, K. Landsteiner, A. Marchon, E. Mottola, P.N.H. Vuong and the referees for helpful discussions. The work of R.L is supported by the Science and Technology Commission of Shanghai Municipality (grant No. 24ZR1450600). The work of D.S. and J.Q. is supported by the EFFORT project, funded through the IRGA program of Université Grenoble Alpes (UGA), and by the Tremplin project from CNRS Physique.

\bibliographystyle{utphys}
\bibliography{bibliography}

\clearpage
\appendix
\setcounter{secnumdepth}{2}
\onecolumngrid

In this Supplementary Material, we present details of various calculations leading to the main results discussed in the text. After writing down our conventions, we detail the computation of the CVE and CSE, and how they contribute to the chiral anomaly. We then explain how the master integrals at finite temperature and electro-chemical potential are computed in curved spacetime.

\section{Conventions and Notations}
\label{app:conventions}

In Lorentzian spacetime $g^L$, the metric signature is $(+---)$. In imaginary-time, the metric becomes Euclidean with $g^E_{tt}=g^L_{tt}$, $g^E_{ti}=ig^L_{ti}$, $g^E_{ij}=-g^L_{ij}$, and the Euclidean Dirac matrices are $\gamma^t_E=i\gamma^t$ and $\gamma^i_E=-\gamma^i$. From Eq.~\eqref{eq:anomCDE} on, the metric and Dirac matrices are Euclidean even though we drop the index $E$. 

The greek indices $\mu,\nu,\rho,\alpha,\beta,\dots$ are the coordinate frame indices and take values in $\{t,x,y,z\}$. The spatial indices $\{x,y,z\}$ are denoted by latin indices starting from $i,j,k,l,\dots$ 
The tangent frame indices are the upper scale latin indices $A,B,C,\dots$ and take values from 0 to 3, whereas the lower scale latin indices (that stop before the index $i$) $a,b,c,\dots$ run from 1 to 3.
\\

\section{Covariant Derivative Expansion}
\label{app:CDE}
In this Appendix, we briefly outline the CDE~\cite{Gaillard:1985uh,Cheyette:1987qz} which is well-suited to deal with curved backgrounds~\cite{Binetruy:1988nx,Alonso:2022ffe,Larue:2023uyv}. We will follow~\cite{Larue:2023uyv} and extend it to imaginary-time. Our goal is to expand
\begin{equation}
\log J[\theta]=\Tr\left[\Big((\slashed\partial\theta)\gamma_5+2im\theta\gamma_5\Big)\frac{1}{i\sD-\gamma^t\sqrt{g_{tt}}\mu_{ec}-m}\right]\;.\nonumber
\end{equation}
We first introduce a basis of plane waves to express the functional trace with a compact and imaginary-time-direction as
\begin{align}
\log J[\theta]=\int_0^{i\beta_0}\dd x^0\int\dd^3 x \frac{1}{i\beta_0}\sum_{n\in\mathds{Z}}\int\frac{\dd^3 q}{(2\pi)^3}e^{-i\omega_n x^0-iq_i x^i}\tr\,\Big((\slashed\partial\theta)\gamma_5+2im\theta\gamma_5\Big)\frac{1}{i\sD-\gamma^t\sqrt{g_{tt}}\mu_{ec}-m}e^{i\omega_n x^0+iq_i x^i}\;,
\end{align}
where $\omega_n=(2n+1)\pi T_0$ and $\partial_\mu q_i=0$. We then define $x^0=it$, and the Jacobian can be expressed as
\begin{equation}
\log J[\theta]=\int_0^{\beta_0}\dd t\int\dd^3 x \frac{1}{\beta_0}\sum_{n\in\mathds{Z}}\int\frac{\dd^3 q}{(2\pi)^3}\tr\,\Big(-(\slashed\partial\theta)_E\gamma_5+2im\theta\gamma_5\Big)\frac{1}{\Delta^{-1}(1-\Delta i\sD_E)}\;,
\end{equation}
where the propagator is $\Delta^{-1}=-(\gamma^t_E\Omega_n+\gamma^i_E q_i+m)$ with $\Omega_n=(2n+1)\pi T_0-i\sqrt{g_{tt}}\mu_{ec}$, and the Euclidean Dirac matrices are defined in App.~\ref{app:conventions}. It can equivalently written as Eq.~\eqref{eq:Delta}. The covariant derivative is also redefined as $\sD_E=-\sD$. Finally, we use
\begin{equation}
\big(A^{-1}(1-AB)\big)^{-1}=\sum_{n\geq0}(AB)^nA\;,
\end{equation}
and $m=0$ to obtain Eq.~\eqref{eq:anomCDE}. In Eqs.~\eqref{eq:Delta} and~\eqref{eq:anomCDE} the subscript $E$ is dropped even though the metric and Dirac matrices are the Euclidean ones.

\section{CVE computation}
\label{app:CVE}

Upon using the Covariant Derivative Expansion the leading order $T$- and $\mu_{ec}$-dependent terms arise at $k=1$ in Eq~\eqref{eq:anomCDE}
\begin{equation}
\left.\langle j^\mu_5\rangle\right|_{k=1}=-\frac{1}{\g}\frac{1}{\beta_0}\sum_{n\in\mathds Z}\int\frac{\dd^3 q}{(2\pi)^3}\tr\,\gamma^\mu\gamma_5\Delta i\sD\Delta=-\frac{1}{\g}\frac{i}{\beta_0}\sum_{n\in\mathds Z}\int\frac{\dd^3 q}{(2\pi)^3}\tr\,\gamma^\mu\gamma_5\left(\Delta \gamma^\nu (D_\nu\Delta)+\Delta\gamma^\nu\Delta (iV_\nu+\omega_\nu))\right)\;.
\label{appeq:j5k=1}
\end{equation}
In Eq.~\eqref{appeq:j5k=1}, the term involving $(D_\mu\Delta)$ arises due to the dependence on $x$ in $\Delta$ via the metric and the electro-chemical potential,  it however vanishes under the Dirac trace by lack of Dirac matrices. In the second term, only the spin-connection provides enough Dirac matrices so that the trace does not vanish, and after performing the integration we obtain
\begin{equation}
\left.\langle j^\mu_5\rangle\right|_{k=1}=-i\frac{1}{\g}\left(\mathcal{I}^2[q^2]\right)_{\alpha\beta}\tr\,\gamma^\mu\gamma_5\gamma^\alpha\gamma^\nu\gamma^\beta\omega_\nu\;.
\label{appeq:j5masterint}
\end{equation}
The master integrals are computed in App.~\ref{app:MasterI}. After performing the Matsubara sums we can identify the contributions from the vacuum, which is the $T$- and $\mu_{ec}$-independent piece. It involves UV divergent integrals which can be regularised using dimensional regularisation. The trace can then be performed using a scheme for $\gamma_5$, and when using the BMHV scheme~\cite{Breitenlohner:1977hr} it is showed in \cite{Larue:2023uyv} that the vacuum contribution cancels in \eqref{appeq:j5masterint}.

The $T$- and $\mu_{ec}$-dependent pieces only involve finite integrals and the Dirac trace can be performed in 4 dimensions to obtain
\begin{equation}
\left.\langle j^\mu_5\rangle\right|_{k=1}=-i\left(\frac{\mu_{ec}^2(x)}{4\pi^2}+\frac{T(x)^2}{12}\right) G_{\alpha\beta}g^{\beta\lambda}\epsilon^{\mu\alpha\rho\sigma}\omega_{\lambda,\rho\sigma}\;,\quad\quad G_{\alpha\beta}\equiv\begin{pmatrix}
 g_{tt} & g_{ti}\\
 g_{ti} & \frac{g_{ti}\,g_{tj}}{g_{tt}}
\end{pmatrix}\;\text{with}\quad G\cdot g^{-1}=\begin{pmatrix}
 1 & 0\\
 \frac{g_{ti}}{g_{tt}} & 0
\end{pmatrix}\;,\nonumber
\end{equation}
where $T(x)=T_0/\sqrt{g_{tt}}$.
We then obtain
\begin{align}
\left.\langle j^\mu_5\rangle\right|_{k=1}=-i\left(\frac{\mu_{ec}^2}{4\pi^2}+\frac{T^2}{12}\right)\left(\epsilon^{\mu t\rho\sigma}+\frac{g_{ti}}{g_{tt}}\epsilon^{\mu i\rho\sigma}\right)\tensor{E}{^A_\rho}\tensor{E}{^B_\sigma}\omega_{t,AB}=i\left(\frac{\mu_{ec}^2}{2\pi^2}+\frac{T^2}{6}\right)\Theta^\mu\;,
\label{eqapp:j5first}
\end{align}
where $\Theta^\mu=-\frac{1}{2}\epsilon^{\mu \nu\rho\sigma}u_\nu u^\lambda\omega_{\lambda,\rho\sigma}$ with $\omega_{\lambda,\rho\sigma}=\tensor{E}{^A_\rho}\tensor{E}{^B_\sigma}\omega_{\lambda,AB}$.

Since the metric and vierbein are independent from the imaginary-time and the expansion is in term is imaginary-time derivative, and we find that $\omega_{t,\rho\sigma}=-g_{\sigma\nu}\Gamma^{\nu}_{t\rho}$.
It is now convenient to introduce the ADM decomposition of the metric (in Euclidean signature)
\begin{equation}
g_{\mu\nu}=\begin{pmatrix}
\alpha^2-v^2 & i v_i \\
i v_i & \sigma_{ij}
\end{pmatrix}\;,\quad 
g^{\mu\nu}=\begin{pmatrix}
\frac{1}{\alpha^2} & -i \frac{v^i}{\alpha^2} \\
-i \frac{v^i}{\alpha^2} & \sigma_{ij}-\frac{v^i v^j}{\alpha^2}
\end{pmatrix}\;,\quad v^i=\sigma^{ij}v_j\;.
\end{equation}
Using the expressions of the Christoffel symbols in ADM decomposition we can work out
\begin{equation}
\epsilon^{\alpha\beta\rho\sigma}\omega_{t,\rho\sigma}=-i\epsilon^{\alpha\beta ij}\partial_i v_j-\epsilon^{\alpha\beta i t}\partial_i (\alpha^2-v^2)=-\epsilon^{\alpha\beta \nu\mu}\partial_{\nu} g_{t\mu}=-\epsilon^{\alpha\beta\rho\sigma}\partial_\rho(\sqrt{g_{tt}}u_\sigma)\;, \label{eqapp:epsilonomega}
\end{equation}
where we used again the imaginary-time-independence of the metric and we recall $u^\mu=\frac{1}{\sqrt{g_{tt}}}\delta^\mu_t$ and $u_{\mu}=\frac{1}{\sqrt{g_{tt}}}g_{t\mu}$.
Inserting \eqref{eqapp:epsilonomega} in \eqref{eqapp:j5first} we obtain for a time-independent metric
\begin{equation}
\Theta^\mu= \Omega^\mu=\frac{1}{2}\epsilon^{\mu\nu\rho\sigma}u_\nu \partial_\rho u_\sigma\;.\nonumber
\end{equation}
Additionally, using $u_t/\sqrt{g_{tt}}=1$ we find that
\begin{equation}
\nabla_\mu\left(\frac{\Theta^\mu}{g_{tt}}\right)=\frac{1}{\sqrt{g}}\partial_\mu\left(\sqrt{g} \frac{\bar\Omega^\mu}{g_{tt}}\right)=\frac{1}{2}\bar\eps^{\mu\nu\rho\sigma}\partial_\mu\left(\frac{u_\nu}{\sqrt{g_{tt}}}\right)\partial_\rho\left(\frac{u_\sigma}{\sqrt{g_{tt}}}\right)=\bar\eps^{i t j k}\partial_i\left(\frac{u_t}{\sqrt{g_{tt}}}\right)\partial_j\left(\frac{u_k}{\sqrt{g_{tt}}}\right)=0\;.\nonumber
\end{equation}
where $\bar\Omega_\mu=\frac{1}{2}\bar\epsilon^{\mu\nu\rho\sigma}u_\nu\partial_\rho u_\sigma$ and $\bar\epsilon^{\mu\nu\rho\sigma}=\sqrt{g}\epsilon^{\mu\nu\rho\sigma}$ is the Levi-Civita tensor density (i.e flat spacetime Levi-Civita tensor) composed of only 0's and $\pm1$'s, i.e $\partial_\lambda\bar\epsilon^{\mu\nu\rho\sigma}=0$. Using $T=T_0/\sqrt{g_{tt}}$, it implies that
\begin{equation}
\nabla_\mu\left(T^2\Theta^\mu\right)=0\;.
\end{equation}
Therefore, the contribution of the CVE to the anomaly in the adiabatic regime is
\begin{equation}
\mathcal{A}_{\mathrm{CVE}}=i\frac{\mu_{ec}}{\pi^2}T\partial_\mu\left(\frac{\mu_{ec}}{T}\right)\Theta^\mu=i\frac{\mu_{ec}}{\pi^2}T\partial_\mu\left(\frac{\mu_{ec}}{T}\right)\Omega^\mu=i\frac{\mu_{ec}}{\pi^2}E_{\mathrm{tot}}\cdot\Omega\;,
\end{equation}
which vanishes at global equilibrium.

\section{CSE computation}
\label{app:CSE}
At $k=2$ in \eqref{eq:anomCDE}, after distributing the covariant derivatives on the propagators and omitting the terms with second derivatives of the fields and thermodynamic variables we obtain
\begin{align}
&\left.\langle j^\mu_5\rangle\right|_{k=2}=\frac{1}{\beta_0}\sum_{n\in\mathds Z}\int\frac{\dd^3q}{(2\pi)^3}\tr\,\gamma^\mu\gamma_5\gamma^\alpha\gamma^\nu\gamma^\beta\gamma^\rho\gamma^\gamma\Bigg(\frac{q_\alpha q_\beta q_\gamma}{(q^2)^3}D_\nu D_\rho\label{eqapp:j5k=2}\\
&+\frac{q_\alpha q_\beta}{(q^2)^2}\left[D_\nu,\frac{q_\gamma}{q^2}\right]D_\rho+\frac{q_\alpha q_\beta}{(q^2)^2}\left[D_\rho,\frac{q_\gamma}{q^2}\right]D_\nu+\frac{q_\alpha q_\gamma}{(q^2)^2}\left[D_\nu,\frac{q_\beta}{q^2}\right]D_\rho\Bigg)\;,\nonumber
\end{align}
where we used $q_\mu\equiv(\Omega_n,q_i)$ as a short-hand.

The term from the first line of \eqref{eqapp:j5k=2} involves only the finite integral $\left(\mathcal{I}^3[q^3]\right)_{\alpha\beta\gamma}$, its vacuum ($T,\mu_{ec}=0$) contribution vanishes identically, since in the vacuum integrals with an odd power of momenta in the numerator vanish. The three terms from the second line of \eqref{eqapp:j5k=2}  involve both $\left(\mathcal{I}^3[q^3]\right)_{\alpha\beta\gamma}$ and $\left(\mathcal{I}^4[q^4]\right)_{\alpha\beta\gamma\delta}$. That latter integral is UV divergent and is regularised in dimensional regularisation (see App.~\ref{app:MasterI}). Note that the second line contributes to an operator of the form $(\partial\mu_{ec}) V_i$, which is not gauge-invariant. However, its divergence $(\partial \mu_{ec})F$ is gauge-invariant and could contribute to the anomaly. Nevertheless, we will see that gauge-variant terms cancel when regularised using the BMHV scheme for $\gamma_5$.

The first line of \eqref{eqapp:j5k=2} reads
\begin{align}
L_1=&\tr\,\gamma^\mu\gamma_5\gamma^\alpha\gamma^\nu\gamma^\beta\gamma^\rho\gamma^\gamma \left(\mathcal{I}^3[q^3]\right)_{\alpha\beta\gamma}\,D_\nu D_\rho\nonumber\\
&=-2\epsilon^{\mu\nu\rho\gamma}g^{\alpha\beta}\frac{1}{\beta_0}\sum_{n\in\mathds Z}\int\frac{\dd^3q}{(2\pi)^3}\frac{1}{(g^{tt}\Omega_n^2+2 g^{ti}q_i\Omega_n+g^{ij}q_i q_j+m^2)^3}\Big[\Omega_n^3\delta^0_\alpha\delta^0_\beta\delta^0_\gamma+q_i\Omega_n^2\left(\delta^0_\alpha\delta^0_\beta\delta^i_\gamma+\delta^0_\alpha\delta^i_\beta\delta^0_\gamma+\delta^i_\alpha\delta^0_\beta\delta^0_\gamma\right)\nonumber\\
&\hspace{6cm}+q_iq_j\Omega_n\left(\delta^0_\alpha\delta^i_\beta\delta^j_\gamma+\delta^i_\alpha\delta^0_\beta\delta^j_\gamma+\delta^i_\alpha\delta^j_\beta\delta^0_\gamma\right)+q_i q_j q_k\delta^i_\alpha\delta^j_\beta\delta^k_\gamma\Big]i\,F_{\nu\rho}\nonumber\;.
\end{align}
We then use the integrals given in App.~\ref{app:MasterI} to obtain
\begin{align}
L_1=&i\frac{\g}{16\pi^2}\mu_{ec}\,\tr\, iF_{\nu\rho}\Big\{g^{tt}g_{tt}\epsilon^{t\mu\nu\rho}+2g_{ti}g^{ti}\epsilon^{t\mu\nu\rho}+g_{ti}g^{tt}\epsilon^{i\mu\nu\rho}+2g_{ij}g^{ti}\epsilon^{j\mu\nu\rho}\nonumber\\
&\hspace{4cm}+g^{ij}g_{ij}\epsilon^{t\mu\nu\rho}+5\frac{g_{ti}}{g_{tt}}\epsilon^{i\mu\nu\rho}+g^{ij}\frac{g_{ti}g_{tj}g_{tk}}{g_{tt}^2}\epsilon^{k\mu\nu\rho}\Big\}\;.\label{eqapp:j5k=2interm}
\end{align}
Using $g^{\mu\rho}g_{\rho\nu}=\delta^\mu_\nu$ and \eqref{eqapp:vierbein} below we find the following identities
\begin{equation}
g_{ti}g^{ti}=1-g_{tt}g^{tt}\;,\quad g_{ij}g^{tj}=-g_{ti}g^{tt}\;,\quad g^{ij}g_{ij}=2+g_{tt}g^{tt}\;,\quad g^{ij}\frac{g_{ti}g_{tj}}{g_{tt}^2}=g^{tt}-\frac{1}{g_{tt}}\;,\label{eqapp:g3+1id}
\end{equation}
which simplify \eqref{eqapp:j5k=2interm} down to
\begin{equation}
L_1=\frac{1}{4\pi^2}\mu_{ec}\left(\epsilon^{t\mu\nu\rho}+\frac{g_{ti}}{g_{tt}}\epsilon^{i\mu\nu\rho}\right)F_{\nu\rho}=\frac{1}{2\pi^2}\mu_{ec} B^\mu\;.\nonumber
\end{equation}
We now turn to the second line of \eqref{eqapp:j5k=2}. Since the integral $\left(\mathcal{I}^4[q^4]\right)_{\alpha\beta\gamma\delta}$ is rather involved in curved spacetime and dealing with the spin-connection $D\supset \omega$ yields lengthy Dirac traces, we restrict the computation for the case of flat spacetime. We also omit higher order contributions in $\mu_{ec}$ (see App.~\eqref{app:MasterI}). We first write explicitly the derivative of the propagators in flat spacetime
\begin{equation}
\left[D_\mu,\frac{q_\nu}{q^2}\right]=\partial_\mu\frac{q_\nu}{q^2}=\frac{i\partial_\mu(\sqrt{g_{tt}}\mu_{ec})}{q^2}\left(-\delta^t_\nu+2g^{\lambda t}\frac{q_\lambda q_\nu}{q^2}\right)\;.\nonumber
\end{equation}
Let us compute the first two terms of the second line of \eqref{eqapp:j5k=2}
\begin{equation}
L_{2,a}^\mu=\frac{1}{\beta_0}\sum_{n\in\mathds Z}\int\frac{\dd^3q}{(2\pi)^3}\tr\,\gamma^\mu\gamma_5\gamma^\alpha\gamma^\nu\gamma^\beta\gamma^\rho\gamma^\gamma\Bigg(\frac{q_\alpha q_\beta}{(q^2)^2}\left[D_\nu,\frac{q_\gamma}{q^2}\right]D_\rho+(\nu\leftrightarrow \rho)\Bigg)\;.
\end{equation}
Using the symmetry of the momenta the trace is simply
\begin{equation}
q_\alpha q_\beta\,\tr\,\gamma^\mu\gamma_5\gamma^\alpha\gamma^\nu\gamma^\beta\gamma^\rho\gamma^\gamma=4q_\alpha q_\beta\,(g^{\alpha\beta}\epsilon^{\gamma\mu\nu\rho}+2g^{\beta\nu}\epsilon^{\gamma\mu\rho\alpha})\;,
\end{equation}
where the first term will not contribute since it is antisymmetric in $\nu\leftrightarrow\rho$. We thus obtain
\begin{equation}
L_{2,a}^\mu=8\epsilon^{\gamma\mu\rho\alpha}g^{\beta\nu}\Big(-i\partial_\nu(\sqrt{g_{tt}}\mu_{ec})\delta^t_\gamma (\mathcal{I}^3[q^2])_{\alpha\beta}D_\rho+(\nu\leftrightarrow\rho)\Big)\;.
\end{equation}
Using the master integrals neglecting higher order contributions in $\mu_{ec}$, we obtain
\begin{equation}
L_{2,a}^\mu=\frac{1}{4\pi^2}\partial_\nu(\sqrt{g_{tt}}\mu_{ec})V_\rho\left(\frac{1}{\bar\epsilon}+\log\left(\frac{\beta}{\mu_{ren}}\right)\right)\left(\epsilon^{t\mu\rho i}\delta^\nu_i+\epsilon^{t\mu\nu i}\delta^\rho_i\right)=0\;,
\end{equation}
since $\epsilon^{t\mu\rho i}\delta^\nu_i=\epsilon^{t\mu\rho i}\delta^\nu_i+\epsilon^{t\mu\rho t}\delta^\nu_t=\epsilon^{t\mu\rho \lambda}\delta^\nu_\lambda=\epsilon^{t\mu\rho \nu}$ and likewise $\epsilon^{t\mu\nu i}\delta^\rho_i=\epsilon^{t\mu\nu \rho}=-\epsilon^{t\mu\rho \nu}$. $\mu_{ren}$ is the renormalisation scale and the spatial dimension is $d=3-\epsilon$.
Finally, we turn to the last term of the second line of \eqref{eqapp:j5k=2}
\begin{align}
L_{2,b}^\mu&=\frac{1}{\beta_0}\sum_{n\in\mathds{Z}}\int\frac{\dd^3q}{(2\pi)^3}\tr\,\gamma^\mu\gamma_5\gamma^\alpha\gamma^\nu\gamma^\beta\gamma^\rho\gamma^\gamma\frac{q_\alpha q_\gamma}{(q^2)^2}\left[D_\nu,\frac{q_\beta}{q^2}\right]iV_\rho\nonumber\\
&=4\left(\epsilon^{\beta\mu\nu\rho}g^{\alpha\gamma}+2\epsilon^{\mu\nu\rho\alpha}g^{\beta\gamma}+2\epsilon^{\beta\mu\rho\alpha}g^{\nu\gamma}-2\epsilon^{\beta\mu\nu\alpha}g^{\rho\gamma}\right)\partial_\nu(\sqrt{g_{tt}}\mu_{ec})V_\rho\left(-\delta^t_\beta(\mathcal{I}^3[q^2])_{\alpha\gamma}+2g^{t\delta}(\mathcal{I}^4[q^4])_{\alpha\beta\gamma\delta}\right)\nonumber\;.
\end{align}
Using the master integrals and $g^{ij}g_{ij}=3-\epsilon$ we obtain
\begin{alignat}{2}
L_{2,b}^\mu&=(\partial_\nu\sqrt{g_{tt}}\mu_{ec})V_\rho
&\Bigg\{&\frac{1}{8\pi^2}\left(\frac{1}{\bar\epsilon}+\log\left(\frac{\beta}{\mu_{ren}}\right)+1\right)\epsilon^{t\mu\nu\rho}\nonumber\\
& &-&\frac{1}{8\pi^2}\left(\frac{1}{\bar\epsilon}+\log\left(\frac{\beta}{\mu_{ren}}\right)\right)\left((3-\epsilon)\epsilon^{t\mu\nu\rho}+2\epsilon^{t\mu\rho i}\delta^\nu_i-2\epsilon^{t\mu\nu i}\delta^\rho_i\right)\nonumber\\
& &-&\frac{1}{8\pi^2}\left(\frac{1}{\bar\epsilon}+\log\left(\frac{\beta}{\mu_{ren}}\right)+\frac{4}{3}\right)\epsilon^{t\mu\nu\rho}-\frac{1}{24\pi^2}\left(\frac{1}{\bar\epsilon}+\log\left(\frac{\beta}{\mu_{ren}}\right)+1\right)(3-\epsilon)\epsilon^{t\mu\nu\rho}\Bigg\}\nonumber\\
&=0\nonumber\;.
\end{alignat}
Finally the contribution from the second line of \eqref{eqapp:j5k=2} vanishes. Let us note that the UV divergences cancel each other and that no renormalising counterterms are involved. Besides, the finite terms only vanish when the contribution of the form $\epsilon/\epsilon$ are properly taken into account.

This shows that, in flat spacetime, while neglecting higher powers of $\mu_{ec}$ and second derivatives of the fields and thermodynamic variables, the $k=2$ contribution to the expectation value of the axial current is gauge invariant and is given by Eq.~\eqref{eq:CSE}. If these hypotheses are relaxed, the computation becomes much more involved, and the sum-integrals are not straightforward to compute.

The divergence of $\langle j^\mu_5\rangle_{k=2}=\frac{\mu_{ec}}{2\pi^2}B^\mu$ can be expressed as
\begin{equation}
\nabla_\mu\langle j^\mu_5\rangle_{k=2}=\frac{1}{2\pi^2}E_{\mathrm{tot}}\cdot B+\frac{\mu_{ec}}{2\pi^2}\sqrt{g_{tt}}\nabla_{\mu}\left(\frac{B^\mu}{\sqrt{g_{tt}}}\right)=\frac{1}{2\pi^2}E_{\mathrm{tot}}\cdot B+\frac{\mu_{ec}}{2\pi^2}\left(\frac{\partial_\mu T}{T}-a_\mu\right)B^\mu\;,
\end{equation}
where $E_{\mathrm{tot}}=T\partial_\mu\frac{\mu_{ec}}{T}$, we used the Schouten identity and $F_{ti}=0$ (since $V_i$ is imaginary-time-independent and $V_t$ was absorbed in $\mu_{ec}$).

Using Bianchi identity, the last term can be recast as
\begin{equation}
\left(\frac{\partial_\mu T}{T}-a_\mu\right)B^\mu=\epsilon^{ijk}F_{jk}\left(\partial_t u_i-u_i\frac{\partial_t T}{T}\right)\;,
\end{equation}
which vanishes since the background fields do not depend on the imaginary-time.

\section{Master integrals}
\label{app:MasterI}
The master integrals at inverse temperature $\beta_0$ and electro-chemical potential $\mu_{ec}$ are
\begin{equation}
\mathcal{I}^k=\frac{1}{\beta_0}\sum_{n\in\mathds{Z}}\int\frac{\dd^3 q}{(2\pi)^3}\frac{1}{(g^{\mu\nu}q_\mu q_\nu+m^2)^k}\;,\quad\quad\left(\mathcal{I}^k[q^{l}]\right)_{\alpha_1\dots\alpha_{l}}=\frac{1}{\beta_0}\sum_{n\in\mathds{Z}}\int\frac{\dd^3 q}{(2\pi)^3}\frac{q_{\alpha_1}\dots q_{\alpha_{l}}}{(g^{\mu\nu}q_\mu q_\nu+m^2)^k}\;,
\end{equation}
where we used $q_\mu\equiv (\Omega_n,q_i)$ as a short-hand. The integrals with momenta on the numerator are fully symmetric in their indices. Note that for $\mu_{ec}=0$, integrals with an odd power of momenta in the numerator always vanish.

To perform the Matsubara sums and integration over 3-momenta, it is convenient to diagonalise the denominator with a first change of variables\footnote{Note that this change of variable may be complex. In practice we assume it is real and then analytically continue to complex values.}
\begin{equation}
p_i=q_i-\frac{g_{ti}}{g_{tt}}\Omega_n\;\Rightarrow
g^{\mu\nu}q_\mu q_\nu=g^{tt}\Omega_n^2+2g^{ti}\Omega_n q_i+g^{ij}q_i q_j=\tilde\Omega_n^2+g^{ij}p_i p_j\;,\label{eq:MasterICHangeVar}
\end{equation}
where $\tilde{\Omega}_n=\Omega_n/\sqrt{g_{tt}}=\pi(2n+1)T(x)-i\mu_{ec}(x)$, $T=1/\sqrt{\beta^2}=T_0/\sqrt{g_{tt}}$, and we used identities from \eqref{eqapp:g3+1id}. This is followed by a second change of variables to diagonalise the 3-momentum part
\begin{equation}
r_a=\tensor{e}{_a^i}p_i\;,\quad \quad p_i=\tensor{e}{^a_i}r_a\;,\quad\quad \frac{1}{\beta_0}\dd^3p=\frac{1}{\beta}\sqrt{g_{tt}}\sqrt{|\det\gamma_{ij}|}\dd^3 r=\frac{1}{\beta}\sqrt{|\det g_{\mu\nu}|}\dd^3r\;,
\end{equation}
where $e_a$ are the 3-dimensional vierbein such that
\begin{equation}
\tensor{e}{^a_i}\tensor{e}{^b_j}\delta_{ab}=\gamma_{ij}\;,\quad\quad \tensor{e}{_a^i}\tensor{e}{_b^j}\delta^{ab}=\gamma^{ij}\;,\quad\quad \gamma^{ij}\equiv g^{ij}\;,\quad\quad \gamma_{ij}=g_{ij}-\frac{g_{ti}g_{tj}}{g_{tt}}\;.\label{eqapp:vierbein}
\end{equation}
After these changes of variables, the denominator is diagonal and any integral with an odd power of $r_a$ in the numerator vanishes.

\subsection{Matsubara sums}

The Matsubara sums are defined as
\begin{equation}
S_n=\frac{1}{\beta}\sum_{k\in\mathds{Z}}\frac{1}{(\tilde\Omega_k^2+E_r^2)^n} \quad\quad \tilde\Omega_k=(2k+1)\pi T(x)-i\mu_{ec}(x)=\Omega_k/\sqrt{g_{tt}}\;,
\end{equation}
with $T(x)=1/\sqrt{\beta^2}=T_0/\sqrt{g_{tt}}$.
They can be computed using
\begin{equation}
S_{n+1}=\frac{(-1)^n}{n! E_r^{2n}}\left[\frac{d^n}{dz^n}S_1(z)\right]_{z=1}\;,\text{ where }\quad S_1(z)=\left.S_1\right|_{E_r\to\sqrt{z}E_r}\;,\quad\quad S_1=\frac{1-\, n(E_r+\mu_{ec})-n(E_r-\mu_{ec})}{2E_r}\;,\label{eqapp:recursionS}
\end{equation}
and we introduced the Fermi-Dirac distribution
\begin{equation}
n(E_r)=\frac{1}{1+e^{\beta(x) E_r}}\;.\nonumber
\end{equation}
Similarly, we define
\begin{equation}
\mathcal{T}_n=\frac{1}{\beta}\sum_{k\in\mathds{Z}}\frac{\tilde\Omega_k}{(\tilde\Omega_k^2+E_r^2)^n}\;,\quad\quad \mathcal{T}_2=-i\frac{\beta}{4E_r}\left(e^{\beta(E_r+\mu_{ec})}n(E_r+\mu_{ec})^2-e^{\beta(E_r-\mu_{ec})}n(E_r-\mu_{ec})^2\right)\;,
\end{equation}
where $\mathcal{T}_2$ is obtained from $S_1$ by
\begin{equation}
\mathcal{T}_2=\frac{1}{2i}\left[\frac{d}{dz} \left.S_1\right|_{\mu\to\mu+z}\right]_{z=0}\;,\nonumber
\end{equation}
and we have the same recursion formula as \eqref{eqapp:recursionS} by replacing every occurrence of $S$ by $\mathcal{T}$ and starting from $\mathcal{T}_2$ instead.

The sums involving higher powers of Matsubara frequencies in the numerator can be dealt with as follows
\begin{equation}
\frac{1}{\beta}\sum_{k\in\mathds{Z}}\frac{\tilde\Omega_k^2}{(\tilde\Omega_k^2+E_r^2)^n}=\frac{1}{\beta}\sum_{k\in\mathds{Z}}\frac{\tilde\Omega_k^2+E_r^2-E_r^2}{(\tilde\Omega_k^2+E_r^2)^n}=S_{n-1}-E_r^2 S_n\;,\quad\quad \frac{1}{\beta}\sum_{k\in\mathds{Z}}\frac{\tilde\Omega_k^3}{(\tilde\Omega_k^2+E_r^2)^n}=\mathcal{T}_{n-1}-E_r^2 \mathcal{T}_n\;.\nonumber
\end{equation}
and so on for higher powers.
\vspace{-5mm}

\subsection{$\left(\mathcal{I}^2[q^2]\right)_{\alpha\beta}$ integral}

Using the change of variables described above and since an odd power in $r$ in the numerator gives a vanishing integral we obtain
\begin{align}
\left(\mathcal{I}^2_\beta[q^2]\right)_{\alpha\beta}&=\g\frac{1}{\beta}\sum_{n\in\mathds Z}\int\frac{\dd^3 r}{(2\pi)^3}\frac{\delta^t_\alpha\delta^t_\beta \Omega_n^2+2\Omega_n^2\frac{g_{ti}}{g_{tt}}(\delta^0_\alpha\delta^i_\beta+\delta^i_\alpha\delta^0_\beta)+\tensor{e}{^a_i}\tensor{e}{^b_j}r_a r_b\delta^i_\alpha\delta^j_\beta+\Omega_n^2\frac{g_{ti}g_{tj}}{g_{tt^2}}\delta^i_\alpha\delta^j_\beta}{(\tilde\Omega_n^2+\delta^{ab}r_a r_b+m^2)^2}\;.\nonumber
\end{align}
Using the symmetry of the integral $\int \dd^3r\, r_a r_b f(r)=\frac{1}{3}\delta_{ab}\int \dd^3r\,r^2 f(r)$, Eq.~\eqref{eqapp:vierbein} and performing the Matsubara sums, we obtain
\begin{align}
\left(\mathcal{I}^2_\beta[q^2]\right)_{\alpha\beta}&=\frac{\g}{2\pi^2}\left(G_{\alpha\beta}\int_0^\infty\dd r\,r^2(S_1-E_r^2 S_2)+\frac{1}{3}\gamma_{ij}\delta^i_\alpha\delta^j_\beta\int_0^\infty\dd r\,r^4 S_2\right)\;\text{ where }\;G_{\alpha\beta}\equiv\begin{pmatrix}
 g_{tt} & g_{ti}\\
 g_{ti} & \frac{g_{ti}\,g_{tj}}{g_{tt}}
\end{pmatrix}\nonumber\;.
\end{align}
Subtracting the vacuum contribution, we find
\begin{align}
\underset{m\to0}{\lim}\left.\left(\mathcal{I}^2_\beta[q^2]\right)_{\alpha\beta}\right|_{T,\mu\text{-dep}}&=\frac{\g}{4\pi^2}\left(\frac{\mu_{ec}^2}{4}+\frac{\pi^2}{12}T^2 \right)\left(G_{\alpha\beta}-\gamma_{ij}\delta^i_\alpha\delta^j_\beta\right)\;,
\end{align}
where we used
\begin{align}
&\int_0^\infty\dd r\,r\,\Big(n(r+\mu_{ec})+n(r-\mu_{ec})\Big)=\frac{\mu_{ec}^2}{2}+\frac{\pi^2}{6}T^2\;,\nonumber\\
&\int_0^\infty\dd r\,r^2 \left(e^{\beta (r+\mu_{ec})}n(r+\mu_{ec})^2+e^{\beta (r-\mu_{ec})}n(r-\mu_{ec})^2\right)=T\left(\mu_{ec}^2+\frac{\pi^2}{3}T^2\right)\;.\nonumber
\end{align}

\subsection{$\left(\mathcal{I}^3[q^3]\right)_{\alpha\beta\gamma}$ integral}

We have
\begin{align}
\left(\mathcal{I}^3[q^3]\right)_{\alpha\beta\gamma}&=\frac{1}{\beta}\sum_{k\in\mathds{Z}}\int\frac{\dd^3q}{(2\pi)^3}\frac{1}{(g^{tt}\Omega_k^2+2 g^{ti}q_i\Omega_k+g^{ij}q_i q_j+m^2)^3}\Big[\Omega_k^3\delta^t_\alpha\delta^t_\beta\delta^t_\gamma+q_i\Omega_k^2\left(\delta^t_\alpha\delta^t_\beta\delta^i_\gamma+\delta^t_\alpha\delta^i_\beta\delta^t_\gamma+\delta^i_\alpha\delta^t_\beta\delta^t_\gamma\right)\nonumber\\
&\hspace{6cm}+q_iq_j\Omega_k\left(\delta^t_\alpha\delta^i_\beta\delta^j_\gamma+\delta^i_\alpha\delta^t_\beta\delta^j_\gamma+\delta^i_\alpha\delta^j_\beta\delta^t_\gamma\right)+q_i q_j q_k\delta^i_\alpha\delta^j_\beta\delta^k_\gamma\Big]\nonumber\;.
\end{align}
Below we compute each term in the massless limit, making use of
\begin{align}
\mathcal{T}_3&=\frac{i\beta}{8 E_r^2}\left(-\frac{1}{2 E_r}e^{\beta(E_r+\mu_{ec})}n(E_r+\mu_{ec})^2-\beta e^{2\beta(E_r+\mu_{ec})}n(E_r+\mu_{ec})^3+\frac{\beta}{2}e^{\beta(E_r+\mu_{ec})}n(E_r+\mu_{ec})^2\right)-(\mu_{ec}\to-\mu_{ec})\nonumber\\
&=-\frac{i\beta}{8 E_r^2}\left(\frac{1}{2 E_r}e^{\beta(E_r+\mu_{ec})}n(E_r+\mu_{ec})^2+\beta \,\mathrm{csch}\left(\beta(E_r+\mu_{ec})\right)^3 \mathrm{sinh}\left(\frac{\beta}{2}\left(E_r+\mu_{ec}\right)\right)^4\right)-(\mu_{ec}\to-\mu_{ec})\;,\nonumber
\end{align}
where csch is the hyperbolic cosecante and sinh is the hyperbolic sinus; under this form it can be straightforwardly integrated in the massless limit
\begin{equation}
\int_0^\infty \dd r\,r^2\mathcal{T}_2=i\frac{\mu_{ec}}{4}\;,\quad\quad\int_0^\infty \dd r\,r^4\mathcal{T}_3=i\frac{3\mu_{ec}}{16}\;.
\end{equation}
The different contributions are
\begin{align}
\frac{1}{\beta}\sum_{k\in\mathds{Z}}\int\frac{\dd^3q}{(2\pi)^3}\frac{\Omega_k^3}{(g^{tt}\Omega_k^2+2 g^{ti}q_i\Omega_k+g^{ij}q_i q_j+m^2)^3}&=\g\frac{1}{\beta}\sum_{k\in\mathds{Z}}\int\frac{\dd^3r}{(2\pi)^3}\frac{g_{tt}^{3/2}\tilde\Omega_k^3}{(\tilde\Omega_k^2+\delta^{ab}r_a r_b+m^2)^3}\nonumber\\
&=\frac{\g g_{tt}^{3/2}}{2\pi^2}\int_0^\infty \dd r \,r^2 \left(\mathcal{T}_2-E_r^2\mathcal{T}_3\right)\underset{m=0}{=}i\frac{\g}{32\pi^2}g_{tt}\,\mu_{ec}\;,\nonumber
\end{align}
\begin{align}
\frac{1}{\beta}\sum_{k\in\mathds{Z}}\int\frac{\dd^3q}{(2\pi)^3}\frac{q_i \Omega_k^2}{(g^{tt}\Omega_k^2+2 g^{ti}q_i\Omega_k+g^{ij}q_i q_j+m^2)^3}&=\g\frac{1}{\beta}\sum_{k\in\mathds{Z}}\int\frac{\dd^3r}{(2\pi)^3}\frac{\Omega_k^2(\tensor{e}{^a_i}r_a+\frac{g_{ti}}{g_{tt}}\Omega_k)}{(\tilde\Omega_k^2+\delta^{ab}r_a r_b+m^2)^3}\nonumber\\
&=\g\frac{g_{ti}}{g_{tt}}g_{tt}^{3/2}\frac{1}{2\pi^2}\int_0^\infty \dd r \,r^2\left(\mathcal{T}_2-E_r^2\mathcal{T}_3\right)\underset{m=0}{=}i\frac{\g}{32\pi^2}g_{ti}\,\mu_{ec}\;,\nonumber
\end{align}
where in the first line the term with an odd power in $r$ in the numerator vanishes.
\begin{align}
\frac{1}{\beta}\sum_{k\in\mathds{Z}}\int\frac{\dd^3q}{(2\pi)^3}\frac{q_i q_j \Omega_k}{(g^{tt}\Omega_k^2+2 g^{ti}q_i\Omega_k+g^{ij}q_i q_j+m^2)^3}&=\g\frac{1}{\beta}\sum_{k\in\mathds{Z}}\int\frac{\dd^3r}{(2\pi)^3}\frac{\Omega_k\left(\tensor{e}{^a_i}\tensor{e}{^b_j}r_a r_b +\frac{g_{ti}g_{tj}}{g_{tt}^2}\right)}{(\tilde\Omega_k^2+\delta^{ab}r_a r_b+m^2)^3}\nonumber\\
&=\frac{\g}{2\pi^2}\left(\frac{g_{tt}^{1/2}}{3}\gamma_{ij}\int_0^\infty \dd r\, r^4\mathcal{T}_3+g_{tt}^{3/2}\frac{g_{ti}g_{tj}}{g_{tt}^2}\int_0^\infty\dd r\,r^2\left(\mathcal{T}_2-E_r^2\mathcal{T}_3\right)\right)\nonumber\\
&\underset{m=0}{=}i\frac{\g}{32\pi^2}\mu_{ec}\left(\gamma_{ij}+\frac{g_{ti}g_{tj}}{g_{tt}}\right)=i\frac{\g}{32\pi^2} g_{ij}\,\mu_{ec}\;.\nonumber
\end{align}
where again the terms with odd powers in $r$ in the numerator vanish and we used $\int \dd^3 r\, r_a r_b f(r)=\frac{1}{3}\delta_{ab}\int \dd^3r\, r^2 f(r)$ and \eqref{eqapp:vierbein}.
Likewise, we find
\begin{align}
&\frac{1}{\beta}\sum_{k\in\mathds{Z}}\int\frac{\dd^3q}{(2\pi)^3}\frac{q_i q_j q_k}{(g^{tt}\Omega_k^2+2 g^{ti}q_i\Omega_k+g^{ij}q_i q_j+m^2)^3}\nonumber\\
&=\g\frac{1}{\beta}\sum_{k\in\mathds{Z}}\int\frac{\dd^3r}{(2\pi)^3}\frac{\frac{g_{ti}g_{tj}g_{tk}}{g_{tt}^3}g_{tt}^{3/2}\tilde\Omega_k+g_{tt}^{1/2}\tilde\Omega_k r_a r_b\left(\frac{g_{ti}}{g_{tt}}\tensor{e}{^a_j}\tensor{e}{^b_k}+\frac{g_{tj}}{g_{tt}}\tensor{e}{^a_i}\tensor{e}{^b_k}+\frac{g_{tk}}{g_{tt}}\tensor{e}{^a_i}\tensor{e}{^b_j}\right)}{(\tilde\Omega_k^2+\delta^{ab}r_a r_b+m^2)^3}\nonumber\\
&=\frac{\g}{2\pi^2}\left(\frac{g_{ti}g_{tj}g_{tk}}{g_{tt}^{3/2}}\int_0^\infty \dd r\,r^2\left(\mathcal{T}_2-E_r^2\mathcal{T}_3\right)+\frac{1}{3g_{tt}^{1/2}}(g_{ti}\gamma_{jk}+g_{tj}\gamma_{ik}+g_{tk}\gamma_{ij})\int_0^\infty \dd r\,r^4\mathcal{T}_3\right)\nonumber\\
&\underset{m=0}{=}i\frac{\g}{32\pi^2}\mu_{ec}\left(g_{ti}\gamma_{jk}+g_{tj}\gamma_{ik}+g_{tk}\gamma_{ij}+\frac{g_{ti}g_{tj}g_{tk}}{g_{tt}^{2}}\right)\;.\nonumber
\end{align}

\subsection{$(\mathcal{I}^4[q^4])_{\alpha\beta\gamma\delta}$ integral}

To keep the computation tractable, we compute this integral in flat spacetime. This integral is UV divergent and we compute in dimensional regualrisation by taking the spatial dimension to be $d=3-\epsilon$. In the massless limit we find
\begin{align}
(\mathcal{I}^4[q^4])_{\alpha\beta\gamma\delta}=&\frac{1}{\beta}\sum_{k\in\mathds{Z}}\int\frac{\dd^dq}{(2\pi)^d}\frac{q_\alpha q_\beta q_\gamma q_\delta}{(\Omega_k^2+\vec q\,^2)^4}\nonumber\\
=&\frac{1}{\beta}\sum_{n\in\mathds{Z}}\int\frac{\dd^dq}{(2\pi)^d}\frac{1}{(\Omega_n^2+\vec q\,^2)^4}\Big\{\Omega_n^4\delta^t_\alpha\delta^t_\beta\delta^t_\gamma\delta^t_\delta+\Omega_n^2 q_i q_j(\delta^t_\alpha\delta^t_\beta\delta^i_\gamma\delta^j_\gamma+\delta^t_\alpha\delta^i_\beta\delta^t_\gamma\delta^j_\gamma+\dots)+q_i q_j q_k q_l\delta^i_\alpha\delta^j_\beta\delta^k_\gamma\delta^l_\delta\Big\}\;,\nonumber
\end{align}
where the dots denote the 4 other possible combinations and $\vec q\,^2=\delta^{ij}q_i q_j$.

To see the divergent behavior as $\eps\rightarrow0$, both integral and sum must be evaluated. This can be achieved by adapting the method in \cite{Bedingham_2001} to massless fermions. By using a Mellin transform, i.e introducing a new integral with respect to a parameter $s$, the computation of the sum and of the momentum-integral become separated and easier to compute. Only the sum is different from the bosonic case~\cite{Bedingham_2001}, and is now
\begin{equation}
    2\sum_{n\in\mathds{N}}\Omega_n^{-s}=2(1-2^{-s})\left(\frac{\pi}{\beta}\right)^{-s}\zeta(s)\;,
\end{equation}
where the factor 2 comes from the separation of negative frequencies before introducing the Mellin transform for even powers of $\Omega_n$ in the numerator.
Finally, the $s$-integral can be computed with the residue theorem. In the massive case, there is an infinity of poles and the result is an expansion in powers of $(\beta m)^k$, whose truncation makes sense in the high-temperature/low-mass limit $T\gg m$. Thus, when considering the massless case the result is much simpler and in the present computation the divergence arises from $\zeta(1+\eps)$. It is not straightforward that this method can be applied for a complex shift in the Matsubara frequencies, we therefore only use it for $\mu_{ec}=0$.

For a vanishing $\mu_{ec}$ each term can be computed to obtain
\begin{equation}
\frac{1}{\beta}\sum_{n\in\mathds{Z}}\int\frac{\dd^dq}{(2\pi)^d}\frac{\Omega_n^4}{(\Omega_n^2+\vec q\,^2)^4}=\frac{1}{64\pi^2}\left(\frac{1}{\bar\epsilon}+\log\left(\frac{\beta}{\mu_{ren}}\right)+\frac{4}{3}\right)+\mathcal{O}(\epsilon)\;,
\end{equation}
where $\mu_{ren}$ is the renormalisation scale, $\bar\epsilon^{-1}=\epsilon^{-1}+(\gamma_E+\log(4/\pi))/2$ with $\gamma_E$ the Euler-Mascheroni constant.
Then
\begin{equation}
\frac{1}{\beta}\sum_{n\in\mathds{Z}}\int\frac{\dd^dq}{(2\pi)^d}\frac{\Omega_n^2 q_i q_j}{(\Omega_n^2+\vec q\,^2)^4}=\frac{\delta_{ij}}{d}\frac{1}{\beta}\sum_{n\in\mathds{Z}}\int\frac{\dd^dq}{(2\pi)^d}\frac{\Omega_n^2 \vec q\,^2}{(\Omega_n^2+\vec q\,^2)^4}=\frac{\delta_{ij}}{192\pi^2}\left(\frac{1}{\bar\epsilon}+\log\left(\frac{\beta}{\mu_{ren}}\right)+1\right)+\mathcal{O}(\epsilon)\;,
\end{equation}
and
\begin{equation}
\frac{1}{\beta}\sum_{n\in\mathds{Z}}\int\frac{\dd^dq}{(2\pi)^d}\frac{ q_i q_j q_k q_l}{(\Omega_n^2+\vec q\,^2)^4}=\frac{\delta_{ijkl}}{d(d+2)}\frac{1}{\beta}\sum_{n\in\mathds{Z}}\int\frac{\dd^dq}{(2\pi)^d}\frac{ (\vec q\,^2)^2}{(\Omega_n^2+\vec q\,^2)^4}=\frac{\delta_{ijkl}}{192\pi^2}\left(\frac{1}{\bar\epsilon}+\log\left(\frac{\beta}{\mu_{ren}}\right)\right)+\mathcal{O}(\epsilon)\;,
\end{equation}
where $\delta_{ijkl}=\delta_{ij}\delta_{kl}+\delta_{ik}\delta_{jl}+\delta_{il}\delta_{jk}$.

\section{Leutwyler and Fujikawa procedures}
\label{app:Leutwyler}

In this Appendix, we detail the derivation of the anomaly following the Leutwyler procedure~\cite{Leutwyler:1984de,Leutwyler:1985ar,Leutwyler:1985em}.

\subsection{Leutwyler method}
The Jacobian from the axial transformation \eqref{eq:Jac} can alternatively be computed following the Fujikawa~\cite{Fujikawa:1979ay} or Leutwyler~\cite{Leutwyler:1984de,Leutwyler:1985ar,Leutwyler:1985em} procedures which yield the same expression of the anomaly. The Leutwyler procedure is outlined in the following. We consider the effective action
\begin{equation}
    \mathcal{W}[A_\mu]=-\log \int_{b.c}\dc\bar\psi\dc\psi e^{-\int_0^{\beta_0}\dd t\int\dd^3x\,(\mathcal{L}-A_\mu j^\mu_5)}=-\log\mathrm{det}_{\beta_0} \mathcal O\;,
\end{equation}
with the Lagrangian given in Eq.~\eqref{eq:Lagr} and $A_\mu$ is introduced as an external source for the axial current $j^\mu_5=\bar\psi\gamma^\mu\gamma_5\psi$. Its only purpose is to extract the expectation value of the axial current, it is not a physical field. After performing the Gaussian integration we obtain the functional determinant of the operator
\begin{equation}
\mathcal{O}=i\gamma^\mu(\nabla_\mu+\omega_\mu+i\mathcal V_\mu+iA_\mu\gamma_5)\;,\quad\quad \mathcal{V}_t=\sqrt{g_{tt}}\mu_{ec}\;, \quad \quad \mathcal V_i=V_i\;.
\end{equation}
The subscripts $\beta_0$ indicate the cyclicity of the time direction in the functional determinants and traces. 

After performing the transformation $\delta_\theta A_\mu=\partial_\mu\theta$ and setting $A_\mu=0$ we obtain
\begin{equation}
\left.\delta_\theta \mathcal W\right|_{A_\mu=0}=\int_0^{\beta_0}\dd t\int\dd^3x \,\theta(x)\langle\nabla j_5\rangle=-\left.\delta_\theta\log\mathrm{det}_{\beta_0}\mathcal O\right|_{A_\mu=0}=-\left.\Tr_{\beta_0}\, (\delta_\theta\mathcal O)\mathcal O^{-1}\right|_{A_\mu=0}\;.
\end{equation}
The functional trace on the right needs to be regularised, this can be achieved using a proper time regularisation
\begin{equation}
\mathcal O^{-1}=-\mathcal O^\dagger\lim_{\Lambda\to\infty}\int_{1/\Lambda^2}^\infty\dd\tau\,f'\left(\tau \mathcal O\mathcal O^\dagger\right)\;, \nonumber
\end{equation}
for $f$ some smooth such that $f(0)=1$ and $\forall n\geq0$, $f^{(n)}(+\infty)=0$, e.g $f(x)=\frac{1}{1+x^2}$. This yields the following expression for the anomaly
\begin{equation}
\mathcal{A}(x)=-\frac{\delta}{\delta\theta(x)}\lim_{\Lambda\to\infty}\left.\Tr_{\beta_0}\,(\delta_\theta \mathcal O)\mathcal O^{-1} f\left(\frac{\mathcal O\mathcal O^\dagger}{\Lambda^2}\right)\right|_{A_\mu=0}=2\frac{\delta}{\delta\theta(x)}\lim_{\Lambda\to\infty}\Tr_{\beta_0}\,\theta\gamma_5\frac{\Lambda^2}{\mathcal D^2+\Lambda^2}\;,\quad\quad \mathcal D=\left.\mathcal O\right|_{A_\mu=0}\;.\label{eqapp:LeutwylerAn}
\end{equation}
More details about the procedure can be found for example in~\cite{Larue:2023tmu}.
The same expression for the anomaly \eqref{eqapp:LeutwylerAn} would be obtained following the Fujikawa procedure. 

The crucial difference with our previous computation is the cutoff $\Lambda$ which is sent to infinity. This cutoff appears as a mass in the propagators, and importantly restores Lorentz invariance, hence killing all dependence on temperature and fluid 4-velocity. We prove it below. As a result, the expression of the anomaly \eqref{eqapp:LeutwylerAn} is the usual axial anomaly except that the gauge field $V_\mu$ is replaced by $\mathcal V_\mu$. We thus obtain
\begin{equation}
\mathcal{A}=\frac{1}{8\pi^2}\mathcal F\mathcal{\tilde F}-\frac{1}{192\pi^2}R\tilde R\;,
\end{equation}
where $\mathcal F$ is the field strength for $\mathcal V$ \eqref{eq:ACSEPontryagin}.

\subsection{Infinite mass limit in the master integrals}
The master integrals that occur in the computation using the Leutwyler method are those defined in App.~\ref{app:MasterI} with the replacement $m\to\Lambda$. After the change of variable described in \eqref{eq:MasterICHangeVar} they take the form\footnote{Contrary to our previous computations, here we keep $\mu_{ec}$ in the gauge field $\mathcal V$ instead of in the Matsubara frequencies $\tilde\omega_n=\pi(2n+\kappa)T(x)$ such that integrals with an odd power of $\tilde\omega_n$ in the numerator vanish.}
\begin{align}
\mathcal{I}=\frac{1}{\beta}\sum_{n\in\mathds{N}}\int\frac{\dd^{3} r}{(2\pi)^{3}}\frac{\tilde\omega_n^{2j}r^{2k}}{(\tilde\omega_n^2+r^2+\Lambda^2)^n}=\Lambda^{3+2k+2j-2n}\sum_{n\in\mathds{N}}\int\frac{\dd^{3} s}{(2\pi)^{3}}\dd s_0 f(n\dd s_0,s_a)\;,
\end{align}
where we changed variables $s_a=r_a/\Lambda$ and defined
\begin{equation}
f(x,s_a)=\frac{s^{2k}\left(2\pi x+\pi\kappa\,\dd s_0\right)^{2j}}{\left(\left(2\pi x+\pi\kappa\,\dd s_0\right)^2+s^2+1\right)^n}\;,\quad\quad \dd s_0=\frac{1}{\beta \Lambda}
\end{equation}
We introduced $\kappa=1$ (resp. 0) for fermionic (resp. bosonic) Matsubara frequencies. At fixed $\beta$ and in the $\Lambda\to\infty$ limit (i.e $\dd s_0\to0$), we identify a Riemann integral and obtain
\begin{equation}
\mathcal{I}=\Lambda^{3+2k+2j-2n}\int_{-\infty}^{+\infty}\dd s_0 \int\frac{\dd^{3} s}{(2\pi)^{3}}\frac{s^{2k}\left(2\pi s_0\right)^{2j}}{\left(\left(2\pi s_0\right)^2+s^2+1\right)^n}=\int\frac{\dd r_0\dd^{3} r}{(2\pi)^{4}}\frac{r^{2k}r_0^{2j}}{\left(r_0^2+r^2+\Lambda^2\right)^n}\;,
\end{equation}
where we changed variables $r_0=2\pi \Lambda s_0$, $r_a=\Lambda s_a$. Firstly, the integral is independent from $\kappa$ in the limit $\Lambda\to\infty$, and secondly we obtain that the result is independent from the temperature. We thus recover the $4$-dimensional integral from a $T=0$ computation. Using the symmetries of the integral, the numerator can be recast as involving only $r_0^2+\delta^{ab}r_a r_b$.

\end{document}